\documentclass[10pt,conference]{IEEEtran}

\usepackage{amsmath}
\usepackage{amssymb}
\usepackage{amsthm}
\usepackage{booktabs}
\usepackage{color}
\usepackage{float}
\usepackage[T1]{fontenc}
\usepackage{graphicx}
\usepackage{subcaption}
\usepackage{hyperref}
\usepackage{ifpdf}
\usepackage[utf8]{inputenc}
\usepackage{keyval}
\usepackage{listings}
\usepackage{makecell}
\usepackage{moresize}
\usepackage{multirow}
\usepackage{nicefrac}
\usepackage{paralist}
\usepackage{rotating}
\usepackage{setspace}
\usepackage{tabularx}
\usepackage{url}
\usepackage{wrapfig}
\usepackage{xcolor}
\usepackage{xspace}

\theoremstyle{definition}

\definecolor{listinggray}{gray}{0.95}
\definecolor{darkgray}{gray}{0.7}
\definecolor{commentgreen}{rgb}{0, 0.4, 0}
\definecolor{darkblue}{rgb}{0, 0, 0.4}
\definecolor{middleblue}{rgb}{0, 0, 0.7}
\definecolor{darkred}{rgb}{0.4, 0, 0}
\definecolor{brown}{rgb}{0.5, 0.5, 0}

\usepackage[normalem]{ulem}
\def\cyanuwave{\bgroup \markoverwith{\lower3.5\p@\hbox{\sixly \textcolor{cyan}{\char58}}}\ULon}
\def\reduwave{\bgroup \markoverwith{\lower3.5\p@\hbox{\sixly \textcolor{red}{\char58}}}\ULon}
\def\blueuwave{\bgroup \markoverwith{\lower3.5\p@\hbox{\sixly \textcolor{blue}{\char58}}}\ULon}
\font\sixly=lasy6 
\makeatother
\newif\ifdraft
\ifdraft

\newcommand{\jhanote}[1]{ {\textcolor{red} { ***shantenu: #1 }}}
\newcommand{\gpnote}[1]{{\textcolor{green} {***giannis: #1}}}
\newcommand{\mtnote}[1]{{\textcolor{orange} {***matteo: #1}}}
\newcommand{\todo}[1]{ {\textcolor{brown} { TODO #1 }}}
\newcommand{\note}[1]{ {\textcolor{magenta} { ***Note: #1 }}}
\else
\newcommand{\jhanote}[1]{}
\newcommand{\gpnote}[1]{}
\newcommand{\mtnote}[1]{}
\newcommand{\todo}[1]{}
\newcommand{\note}[1]{}
\fi

\newcommand{\entk}{EnTK\xspace}

\floatstyle{ruled}
\newfloat{scheme}{htbp}{lok}
\floatname{scheme}{Scheme}

\lstdefinestyle{myListing}{
    frame=single,
    backgroundcolor=\color{listinggray},
    language=C,
    basicstyle=\ttfamily \footnotesize,
    breakautoindent=true,
    breaklines=true
    tabsize=2,
    captionpos=b,
    aboveskip=0em,
    belowskip=-2em,
}

\lstdefinestyle{myPythonListing}{
    frame=single,
    backgroundcolor=\color{listinggray},
    language=Python,
    basicstyle=\ttfamily \scriptsize,
    breakautoindent=true,
    breaklines=true
    tabsize=2,
    captionpos=b,
}

\ifpdf
\DeclareGraphicsExtensions{.pdf, .jpg, .tif}
\else
\DeclareGraphicsExtensions{.ps,  .eps, .jpg}
\fi

\usepackage{listings}
\usepackage{paralist}
\lstnewenvironment{code}[1][]%
{
    \noindent
    \minipage{1.0 \linewidth}
    \vspace{0.5\baselineskip}
    \lstset{
        language=Python,
        frame=single,
        captionpos=b,
        stringstyle=\ttfamily,
        basicstyle=\scriptsize\ttfamily,
        showstringspaces=false,#1}
}
{\endminipage}
\begin{document}
    
\title{Workflow Design Analysis for High Resolution Satellite Image Analysis}

\author{
	
	\IEEEauthorblockN{
		Ioannis Paraskevakos\IEEEauthorrefmark{1},
		Matteo Turilli\IEEEauthorrefmark{1},
		Bento Collares Gonçalves\IEEEauthorrefmark{2},
		Heather J. Lynch\IEEEauthorrefmark{2},
		Shantenu Jha\IEEEauthorrefmark{1}\IEEEauthorrefmark{3}\IEEEauthorrefmark{4}}
	
	\IEEEauthorblockA{\IEEEauthorrefmark{1}Department of Electrical and Computer Engineering, Rutgers University, Piscataway, New Jersey 08854}
	\IEEEauthorblockA{
		\IEEEauthorrefmark{2}Department of Ecology and Evolution, Stony Brook, NY USA 11777}
	\IEEEauthorblockA{
		\IEEEauthorrefmark{2}Brookhaven National Laboratory}

	\IEEEauthorblockA{
		\IEEEauthorrefmark{3}Corresponding Author}
	
}

\maketitle
\begin{abstract}
    Ecological sciences are using imagery from a variety of sources to
    monitor and survey populations and ecosystems. Very High Resolution (VHR)
    satellite imagery provide an effective dataset for large scale surveys.
    Convolutional Neural Networks have successfully been employed to analyze
    such imagery and detect large animals. As the datasets increase in
    volume, O(TB), and number of images, O(1k), utilizing High Performance
    Computing (HPC) resources becomes necessary. In this paper, we
    investigate a task-parallel data-driven workflows design to support 
    imagery analysis pipelines with heterogeneous tasks on HPC. We analyze 
    the capabilities of each design when processing a dataset of 3,000 VHR 
    satellite images for a total of 4~TB. We experimentally model the 
    execution time of the tasks of the image processing pipeline. We perform 
    experiments to characterize the resource utilization, total time to 
    completion, and overheads of each design. Based on the model, overhead 
    and utilization analysis, we show which design approach to is best suited 
    in scientific pipelines with similar characteristics.
\end{abstract}


\section{Introduction}\label{sec:intro}

A growing number of scientific domains are adopting workflows that use
multiple analysis algorithms to process a large number of images. The volume
and scale of data processing justifies the use of parallelism, tailored
programming models and high performance computing (HPC) resources. While
these features create a large design space, the lack of architectural and
performance analyses makes it difficult to chose among functionally
equivalent implementations.

In this paper we focus on the design of computing frameworks that support the
execution of heterogeneous tasks on HPC resources to process large imagery
datasets. These tasks may require one or more CPUs and GPUs, implement
diverse functionalities and execute for different amounts of time. Typically,
tasks have data dependences and are therefore organized into workflows. Due
to task heterogeneity, executing workflows poses the challenges of effective
scheduling, correct resource binding and efficient data management. HPC
infrastructures exacerbate these challenges by privileging the execution of
single, long-running jobs.

From a design perspective, a promising approach to address these challenges
is isolating tasks from execution management. Tasks are assumed to be
self-contained programs which are executed in the operating system (OS)
environment of HPC compute nodes. Programs implement the domain-specific
functionalities required by use cases while computing frameworks implement
resource acquisition, task scheduling, resource binding, and data management.

Compared to approaches in which tasks are functions or methods, a
program-based approach offers several benefits as, for example, simplified
implementation of execution management, support of general purpose
programming models, and separate programming of management and
domain-specific functionalities. Nonetheless, program-based designs also
impose performance limitations, including OS-mediated intertask communication
and task spawning overheads, as programs execute as OS processes and do not
share a memory space.

Due to their performance limitations, program-based designs of computing
frameworks are best suited to execute compute-intense workflows in which each
task requires a certain amount of parallelism and runs from several minutes
to hours. The emergence of workflows that require heterogeneous,
compute-intense tasks to process large amount of data is pushing the
boundaries of program-based designs, especially when scale requirements
suggest the use of modern HPC infrastructures with large number of CPUs/GPUs
and dedicated data facilities.

We use a paradigmatic use case from the polar science domain to evaluate
three alternative program-based designs, and experimentally characterize and
compare their performance. Our use case requires us to analyze satellite
images of Antarctica to detect pack-ice seals taken across a whole calendar
year. The resulting dataset consists of $3,097$ images for a total of
$\approx4$~TB. The use case requires us to repeatedly process these images,
running both CPUs and GPUs code that exchange several GB of data. The first
design uses a pipeline to independently process each image, while the second
and third designs use the same pipeline to process a series of images with
differences in how images are bound to available compute nodes.

This paper offers three main contributions: 
\begin{inparaenum}[(1)]
    \item a precise indication of how to further the implementation of our 
    workflow engine so to support the class of use cases we considered while 
    minimizing workflow time to completion and maximizing resource 
    utilization;
    \item specific design guidelines for supporting data-driven,
    compute-intense workflows on high-performance computing resources with a
    task-based computing framework; and
    \item an experiment-based methodology to compare design performance of
    alternative designs that does not depend on the considered use case and
    computing framework.
\end{inparaenum}

The paper is organized as follows. \S\ref{sec:ucase} presents the use case in
more detail and discuses its computational requirements as well as the
individual stages of the pipeline. \S\ref{sec:related} provides a survey of
the state of the art and~\S\ref{sec:design} discusses the three designs in
detail. \S\ref{sec:experiments} presents our performance evaluation,
discussing the results of our experiments and~\S\ref{sec:conclusion}
concludes the paper.

\section{Related Work}\label{sec:related}
Several tools and frameworks are available for image analysis based on
diverse designs and programming paradigms, and implemented for specific
resources. Numerous image analytics frameworks for medical, astronomical, and
other domain specific imagery provide MapReduce implementations.
MaReIA~\cite{vo2018mareia}, built for medical image analysis, is based on
Hadoop and Spark~\cite{zaharia2010spark}. Kira~\cite{zhang2016kira}, built
for astronomical image analysis, is also built on top of Spark and pySpark,
allowing users to define custom analysis applications. Further,
Ref.~\cite{yan2014large} proposes a Hadoop-based cloud Platform as a Service,
utilizing Hadoop's streaming capabilities to reduce filesystem reads and
writes. These frameworks support clouds and/or commodity clusters for
execution.

BIGS~\cite{ramos2012bigs} is a framework for image processing and analysis.
BIGS is based on the master-worker model and supports heterogeneous
resources, such as clouds, grids and clusters. BIGS deploys a number of
workers to resources, which query its scheduler for jobs. When a worker can
satisfy the data dependencies of a job it becomes responsible to execute it.
BIGS workers can be deployed on any type of supported resource. The user is
responsible to define the input, processing pipeline, and launch BIGS
workers. As soon as a worker is available execution starts. In addition,
BISGS offers a diverse set of APIs for developers. BIGS approach is very
close to Design 1 we described in~\S\ref{des1}.

LandLab~\cite{hobley2017creative} is a framework for building, coupling and
exploring two-dimensional numerical models for Earth-surface dynamics.
LandLab provides a library of processing constructs. Each construct is a
numerical representation of a geological process. Multiple components are
used together allowing the simulation of multiple processes acting on a grid.
The design of each component is intended to work in a plug-and-play fashion.
Components couple simply and quickly. Parallelizing Landlab components is 
left to the developer.

Image analysis libraries, frameworks and applications have been proposed for
HPC resources. PIMA(GE)\textsuperscript{2} Library~\cite{galizia2015mpicuda}
provides a low-level API for parallel image processing using MPI and CUDA.
SIBIA~\cite{gholami2017framework} is a framework for coupling biophysical
models with medical image analysis, and provides parallel computational
kernels through MPI and vectorization. Ref.~\cite{huo2018towards} proposes a
scalable medical image analysis service. This service uses
DAX~\cite{damon2017dax} as an engine to create and execute image analysis
pipelines. Tomosaic~\cite{vescovi2018tomosaic} is a Python framework, used
for medical imaging, employing MPI4py to parallelize different parts of the
workflow.

Petruzza et al.~\cite{petruzza2017isavs} describe a scalable image analysis
library. Their approach defines pipelines as data-flow graphs, with user
defined functions as tasks. Charm++ is used as the workflow management layer,
by abstracting the execution level details, allowing execution on local
workstations and HPC resources. Teodoro et
al.~\cite{teodoro2013highthroughput} define a master-worker framework
supporting image analysis pipelines on heterogeneous resources. The user
defines an abstract dataflow and the framework is responsible to schedule
tasks on CPU or GPUs. Data communication and coordination is done via MPI.
Ref.~\cite{grunzke2017seamless} proposes the use of the
UNICORE~\cite{benedyczak2016unicore} to define image analysis workflows on
HPCs.

Our approach proposes designs for image analysis pipelines that are domain
independent. In addition, the workflow and runtime systems we use allow 
execution on multiple HPC resources with no change in our approach. 
Furthermore, parallelization is inferred in one of the proposed designs, 
allowing correct execution regardless of the multi-core or multi-GPU 
capabilities of the used resource.

All the above, except Ref.~\cite{zhang2016kira}, focus on characterizing the
performance of the proposed solution. Ref.~\cite{zhang2016kira} compares
different implementations, one with Spark, one with pySpark, and an MPI
C-based implementation. This comparison is based on the weak and strong
scaling properties of the approaches. Our approach offers a well-defined
methodology to compare different designs for task-based and data-driven
pipelines with heterogeneous tasks.

\section{Satellite Imagery Analysis Application}\label{sec:ucase}
Imagery employed by ecologists as a tool to survey populations and ecosystems
come from a wide range of sensors, e.g., camera-trap
surveys~\cite{karanth1995estimating} and aerial imagery
transects~\cite{western2009impact}. However, most traditional methods can be
prohibitively labor-intensive when employed at large scales or in remote
regions. Very High Resolution (VHR) satellite imagery provides an effective
alternative to perform large scale surveys at locations with poor
accessibility such as surveying Antarctic fauna~\cite{lynch2012detection}. To
take full advantage from increasingly large VHR imagery, and reach the
spatial and temporal breadths required to answer ecological questions, it is
paramount to automate image processing and labeling.

Convolutional Neural Networks (CNN) represent the state-of-the-art for nearly
every computer vision routine. For instance, ecologists have successfully
employed CNNs to detect large mammals in airborne
imagery~\cite{kellenberger2018detecting,polzounov2016right}, and camera-trap
survey imagery~\cite{norouzzadeh2018automatically}. We use a Convolutional
Neural Network (CNN) to survey Antarctic pack-ice seals in VHR imagery.
Pack-ice seals are a main component of the Antarctic food
web~\cite{fabra2008convention}: estimating the size and trends of their
populations is key to understanding how the Southern Ocean ecosystem copes
with climate change~\cite{hillebrand2018climate} and
fisheries~\cite{reid2019climate}.

For this use case, we process WorldView 3 (WV03) panchromatic imagery as
provided by DigitalGlobe Inc. This dataset has the highest available
resolution for commercial satellite imagery. We refrain from using imagery
from other sensors because pack-ice seals are not clearly visible at lower
resolutions. For our CNN architecture, we use a U-Net~\cite{ronneberger2015u}
variant that counts seals with an added regression branch and locates them
using a seal intensity heat map. To train our CNN, we use a training set with
$88k$ hand-labeled images, cropped from WV03 imagery, where every image has a
correspondent seal count and a class label (i.e., seal vs. non-seal). For
hyper-parameter search, we train CNN variants for 75 epochs (i.e., 75
complete runs through the training set) using an Adam
optimizer~\cite{kingma2014adam} with a learning rate of $10^{-3}$ and tested
against a validation set.

We use the best performing model on an archive of over $3,097$ WV03 images,
with a total dataset size of $\approx4$~TB. Due to limitations on GPU memory,
it is necessary to tile WV03 images into smaller patches before sending input
imagery through the seal detection CNN. Taking tiled imagery as input, the
CNN outputs the latitude and longitude of each detected seal. While the raw
model output still requires statistical treatment, such `mock-run' emulates
the scale necessary to perform a comprehensive pack-ice seal census. We order
the tiling and seal detection stages into a pipeline that can be re-run
whenever new imagery is obtained. This allows domain scientists to create
seal abundance time series that can aid in Southern Ocean monitoring.

\section{Workflow Design and Implementation}\label{sec:design}

Computationally, the use case described in~\S\ref{sec:ucase} presents three
main challenges: heterogeneity, scale and repeatability. The images of the
use case dataset vary in size with a wide distribution. Each image requires
tiling, per tile counting of seal populations and result aggregation across
tiles. Tiling is memory intensive while counting is computationally
intensive, requiring CPU and GPU implementations, respectively. Whenever the
image dataset is updated, it needs to be reprocessed.

We address these challenges by codifying image analysis into a workflow. We
then execute this workflow on HPC resources, leveraging the concurrency,
storage systems and compute speed they offer to reduce time to completion.
Typically, this type of workflow consists of a sequence (i.e., pipeline) of
tasks, each performing part of the end-to-end analysis on one or more images.
The implementation of this workflow varies, depending on the API of the
chosen workflow system and its supported runtime system. Here, we compare two
common designs: one in which each image is processed independently by a
dedicated pipeline; the other in which a pipeline processes multiple images.

Note that both designs separate the functionalities required to process each 
image from the functionalities used to coordinate the processing of multiple 
images. This is consistent with moving away from vertical, end-to-end  
single-point solutions, favoring designs and implementations that satisfy 
multiple use cases, possibly across diverse domains. Accordingly, the two 
designs we consider, implement, and characterize use two tasks (i.e., 
programs) to implement the tiling and counting functionalities required by 
the use case.

Both designs are functionally equivalent, in that they both enable the
analysis of the given image dataset. Nonetheless, each design leads to
different amounts of concurrency, resource utilization and overheads,
depending on data/compute affinity, scheduling algorithm, and coordination
between CPU and GPU computations. Based on a performance analysis, it will be
possible to know which design entails the best tradeoffs for common metrics
as total execution time or resource utilization.

Consistent with HPC resources currently available for research and our use 
case, we make three architectural assumptions: 
\begin{inparaenum}[(1)]
    \item each compute node has $c$ CPUs;
    \item each compute node has $g$ GPUs where $g \le c$; and
    \item each compute node has enough memory to enable concurrent execute of
    a certain number of tasks.
\end{inparaenum}
As a result, at any given point in time there are $C = n\times c$ CPUs and $G
= n\times$ GPUs available, where $n$ is the number of compute nodes.

\begin{figure*}[ht!]
    \centering
    \begin{subfigure}[b]{0.32\textwidth}
        \includegraphics[width=\linewidth]{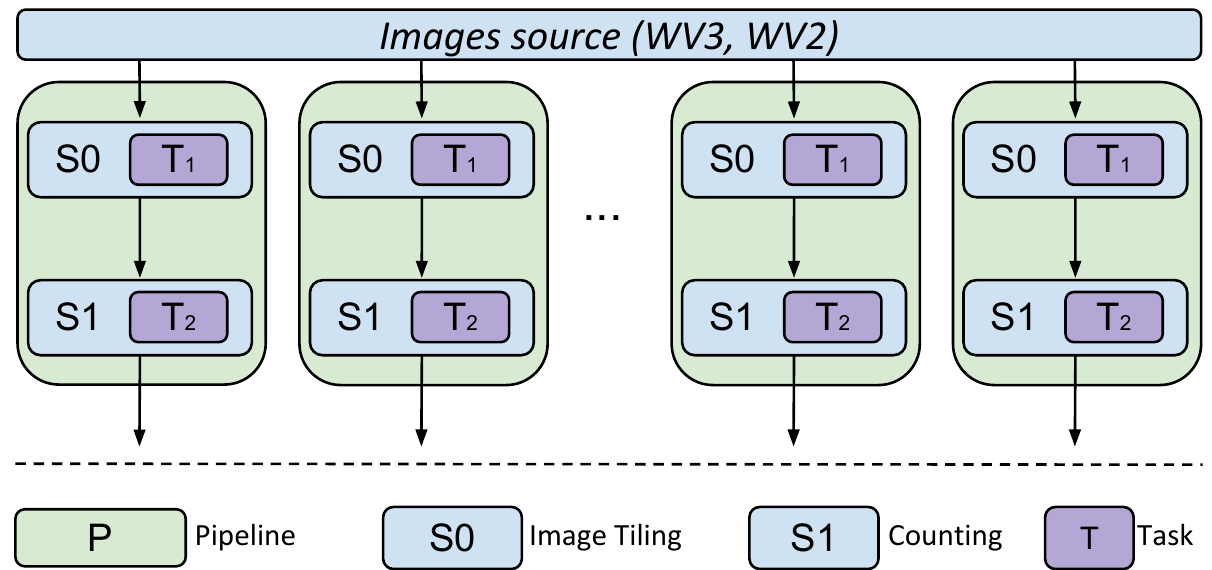}
        \caption{}
		\label{fig:seals_design1}
    \end{subfigure}%
    ~ 
    \begin{subfigure}[b]{0.32\textwidth}
        \includegraphics[width=\linewidth]{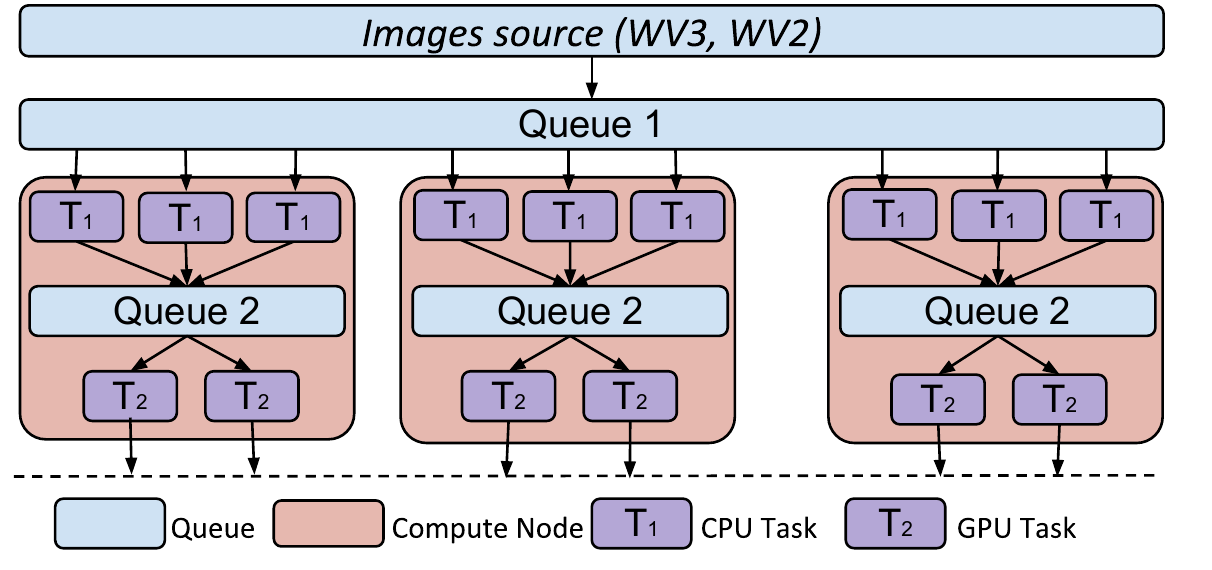}
        \caption{}\label{fig:seals_design2}
    \end{subfigure}%
    ~ 
    \begin{subfigure}[b]{0.32\textwidth}
        \includegraphics[width=\linewidth]{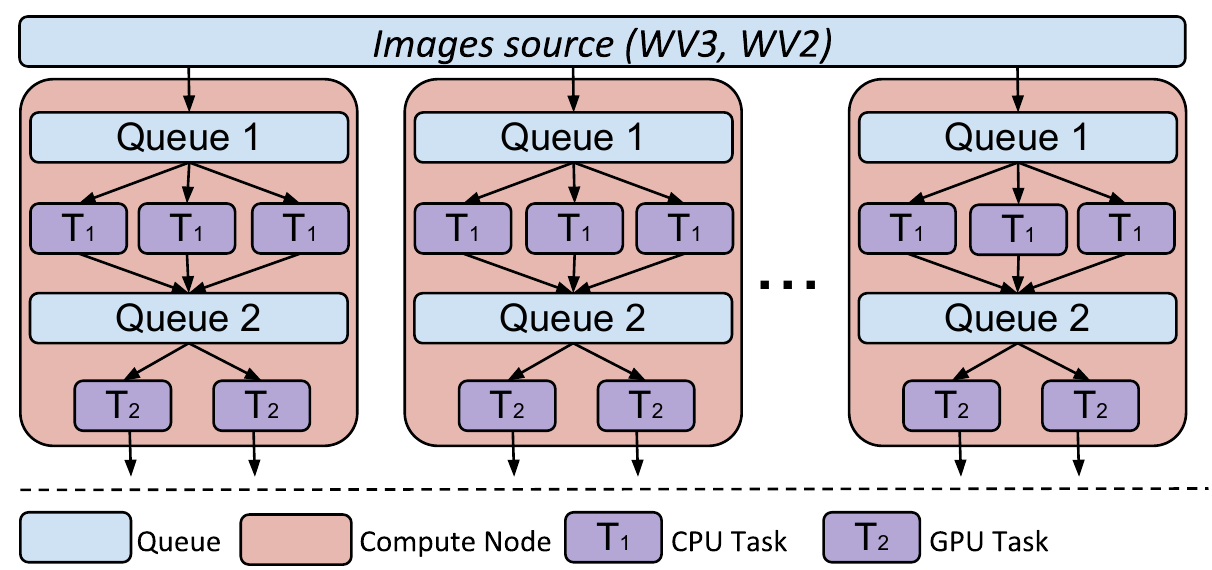}
		\caption{}\label{fig:seals_design3}
    \end{subfigure}
    \caption{Design approaches to implement the workflow required for the
    Seals use case.~\ref{fig:seals_design1}--\textbf{Design 1}: Pipeline,
    stage and task based design.~\ref{fig:seals_design2}--\textbf{Design 2}:
    Queue based design with a single queue for all the tiling
    tasks.~\ref{fig:seals_design3}--\textbf{Design 2.A}: Queue based design
    with multiple queues for the tiling tasks.}\label{fig:designs}
\end{figure*}

\subsection{Design 1: One Image per Pipeline}
\label{ssec:approach1}\label{des1}


We specify the workflow for counting the number of seals in a set of images
as a set of pipelines. Each pipeline is composed of two stages, each with one
type of task. The task of the first stage gets an image as input and
generates tiles of that image based on a selected tile size as output. The
task of the second stage gets the generated tiles as input, counts the number
of seals found in each tile and outputs the aggregated result for the whole
image.

Formally, we define two types of tasks:
\begin{itemize}
    \item $T_{1} = <I, f_{I}, t>$, where $I$ is an image, $f_{I}$ is a tiling 
    function and $t$ is a set of tiles that correspond to $I$.
    \item $T_{2} = <t, f_{A}, S>$, where $f_{A}$ is a function that counts  
    seals from a set of tiles and $S$ is the number of seals.
\end{itemize}

Tiling in $T_{1}$ is implemented with OpenCV~\cite{bradski2000opencv} and
Rasterio~\cite{gillies2013rasterio} in Python. Rasterio allows us to open and
convert a GeoTIFF WV3 image to an array. The array is then partitioned to
sub\-arrays based on a user-specified scaling factor. Each sub\-array is
converted to an compressed image via OpenCV routines and saved to the
filesystem.

Seal counting in $T_{2}$ is performed via a Convolutional Neural Network
(CNN) implemented with PyTorch~\cite{paszke2017automatic}. The CNN counts
the number of seals for each tile of an input image. When all tiles are
processed, the coordinates of the tiles are converted to the geological
coordinates of the image and saved in a file, along with the number of
counted seals.

Both tiling and seal counting implementations are invariant between the
designs we consider. This is consistent with the designs being task-based,
i.e., each task exclusively encapsulates the capabilities required to perform
a specific operation over an image or tile. Thus, tasks are independent
from the capabilities required to coordinate their execution, whether each
task processes a single image or sequence of images.

We implemented Design~1 via \entk, a workflow engine which exposes an API
based on pipelines, stages, and tasks~\cite{balasubramanian2018harnessing}.
The user can define a set of pipelines, where each pipeline has a sequence of
stages, and each stage has a set of tasks. Stages are executed respecting
their order in the pipeline while the tasks in each stage can execute
concurrently, depending on resource availability.

For our use case, \entk{} has three main advantages compared to other
workflow engines:
\begin{inparaenum}[(1)]
    \item it exposes pipelines and tasks as first-order abstractions
    implemented in Python;
    \item it is specifically designed for concurrent management of up to
    $10^5$ pipelines; and
    \item it supports RADICAL-Pilot, a pilot-based runtime system designed to
    execute heterogeneous bag of tasks on HPC
    machines~\cite{merzky2018using}.
\end{inparaenum} 
Together, these features address the challenges of heterogeneity, scale and
repeatability: users can encode multiple pipelines, each with different types
of tasks, executing them at scale on HPC machines without explicitly coding
parallelism and resource management.

When implemented in \entk, the use case workflow maps to a set of pipelines,
each with two stages $St_{1}$, $St_{2}$. Each stage has one task
$T_{1}$ and $T_{2}$ respectively. The pipeline is defined as 
$P = (St_{1},St_{2})$. For our use case the workflow consists of $N$ 
pipelines, where $N$ is the number of images.

Figure~\ref{fig:seals_design1} shows the workflow. For each pipeline, \entk{}
submits the task of stage $St_{1}$ to the runtime system (RTS). As soon as
the task finishes, the task of stage $St_{2}$ is submitted for execution.
This design allows concurrent execution of pipelines and, as a result, 
concurrent analysis of images, one by each pipeline. Since pipelines
execute independently and concurrently, there are instances where $St_{1}$
of a pipeline executes at the same time as $St_{2}$ of another pipeline.

Design~1 has the potential to increase utilization of available resources as
each compute node of the target HPC machine has multiple CPUs and GPUs.
Importantly, computing concurrency comes with the price of multiple reading
and writing to the filesystem on which the dataset is stored. This can cause
I/O bottlenecks, especially if each task of each pipeline reads from and
writes to the same filesystem, possibly over a network connection.

We used a tagged scheduler for \entk's RTS to avoid I~/O bottlenecks. This
scheduler schedules $T_{1}$ of each pipeline on the first available compute
node, and guarantees that the respective $T_{2}$ is scheduled on the same
compute node. As a result, compute/data affinity is guaranteed among
co-located $T_{1}$ and $T_{2}$. While this design avoids I~/O bottlenecks, it
may reduce concurrency when the performance of the compute nodes and/or the
tasks is heterogeneous: $T_{2}$ may have to wait to execute on a specific
compute node while another node is free.

\subsection{Design 2: Multiple images per pipeline}\label{ssec:approach2}

Design~2 implements a queue-based design. We introduce two tasks $T_{1}$ and
$T_{2}$ as defined in~\S\ref{ssec:approach1}. In contrast to Design 1, these
tasks are bootstrapped once and then executed for as long as resources are
available, processing input images at the rate taken to process each image.
The number of concurrent $T_{1}$ and $T_{2}$ depends on available resources,
including CPUs, GPUs, and RAM.

For the implementation of Design~2, we do not need \entk{}, as we submit a 
bag of $T_{1}$ and $T_{2}$ tasks via the RADICAL\nobreakdash-Pilot RTS, and 
manage the data movement between tasks via queues. As shown in 
Fig.~\ref{fig:seals_design2}, Design 2 uses one queue (Queue 1) for the 
dataset, and another queue (Queue 2) for each compute node. For each 
compute node, each $T_{1}$ pulls an image from Queue 1, tiles that image and 
then queues the resulting tiles to Queue 2. The first available $T_{2}$ on 
that compute node, pulls those tiles from Queue 2, and counts the seals.

To communicate data and control signals between queues and tasks, we defined
a communication protocol with three entities: Sender, Receiver, and Queue.
Sender connects to Queue and pushes data. When done, Sender informs Queue and
disconnects. Receiver connects to Queue and pulls data. If there are no data
in Queue but Sender is connected, Receiver pulls a ``wait'' message, waits,
and pulls again after a second. When there are no data in Queue or Sender is
not connected to Queue, Receiver pulls an ``empty'' message, upon which it
disconnects and terminates. This ensures that tasks are executing, even if
starving, and that all tasks are gracefully terminating when all images are
processed.

Note that Design~2 load balances $T_{1}$ tasks across compute nodes but 
balances $T_{2}$ tasks only within each node. For example, suppose that
$T_{1}$ on compute node $A$ runs two times faster than $T_{1}$ on compute
node $B$. Since both tasks are pulling images from the same queue, $T_{1}$ of
$A$ will process twice as many images as $T_{1}$ of $B$. Both $T_{1}$ of $A$
and $B$ will execute for around the same amount of time until Queue 1 is
empty, but Queue 2 of $A$ will be twice as large as Queue 2 of $B$. $T_{2}$
tasks executing on $B$ will process half as many images as $T_{2}$ tasks on
$A$, possibly running for a shorter period of time, depending on the
time taken to process each image.

In principle, Design~2 can be modified to load balance also across Queue 2
but in practice, as discussed in~\S\ref{ssec:approach1}, this would produce
I/O bottlenecks: Load balancing across $T_{2}$ tasks would require for all
tiles produced by $T_{1}$ tasks to be written to and read  from a filesystem
shared across multiple compute nodes. Keeping Queue 2 local to each compute
node enables using the filesystem local to each compute node.

\subsubsection{Design 2.A: Uniform image dataset per pipeline}
\label{sssec:approach2a}

The lack of load balancing of $T_{2}$ tasks in Design 2 can be mitigated by
introducing a queue in each node from where $T_{1}$ tasks pull images. This
allows early binding of images to compute nodes, i.e., deciding the
distribution of images per node before executing $T_{1}$ and $T_{2}$. As a
result, the execution can be load balanced among all available nodes,
depending on the correlation between image properties and image execution
time.

Figure~\ref{fig:seals_design3} shows variation 2.A of Design~2. The early
binding of images to compute nodes introduces an overhead compared to using
late binding via a single queue as in Design 2. Nonetheless, depending on the
strength of the correlation between image properties and execution time,
design 2.A offers the opportunity to improve resource utilization. While in
Design 2 some node may end up waiting for another node to process a much
larger Queue 2, in design 2.A this is avoided by guaranteeing that each
compute node has an analogous payload to process.

\section{Experiments and Discussion}\label{sec:experiments}

We executed three experiments using GPU compute nodes of the XSEDE Bridges
supercomputer. These nodes offer 32 cores, 128 GB of RAM and two P100 Tesla
GPUs. We stored the dataset of the experiments and the output files on XSEDE
Pylon5 Lustre filesystem. We stored the tiles produced by the tailing tasks
on the local filesystem of the compute nodes. This way, we avoided creating a
performance bottleneck by performing millions of reads and writes of
$\approx700$~KB on Pylon5. We submitted jobs requesting 4 compute nodes to
keep the average queue time within a couple of days. Requesting more nodes
produced queue times in excess of a week.

The experiments dataset consists of $3,097$ images, ranging from $50$ to
$2,770$~MB for a total of $\approx4$~TB of data. The images size follows a
normal distribution with a mean value of $1,304.85$~MB and standard deviation
of $512.68$~MB.

For Design~1, 2 and 2.A described in~\S\ref{sec:design}, Experiment~1 models
the execution time of the two tasks of our use case as a function of the
image size (the only property of the images for which we found a correlation
with execution time); Experiment~2 measures the total resource utilization of
each design; and Experiment~3 characterizes the overheads of the middleware
implementing each design. Together, these experiments enable performance
comparison across designs, allowing us to draw conclusions about the
performance of heterogeneous task-based execution of data-driven workflows on
HPC resources.

\subsection{Experiment~1: Design~1 Tasks Execution Time}
\label{ssec:des1analysis}

\begin{figure}[ht!]
    \centering
    \includegraphics[width=0.45\textwidth]{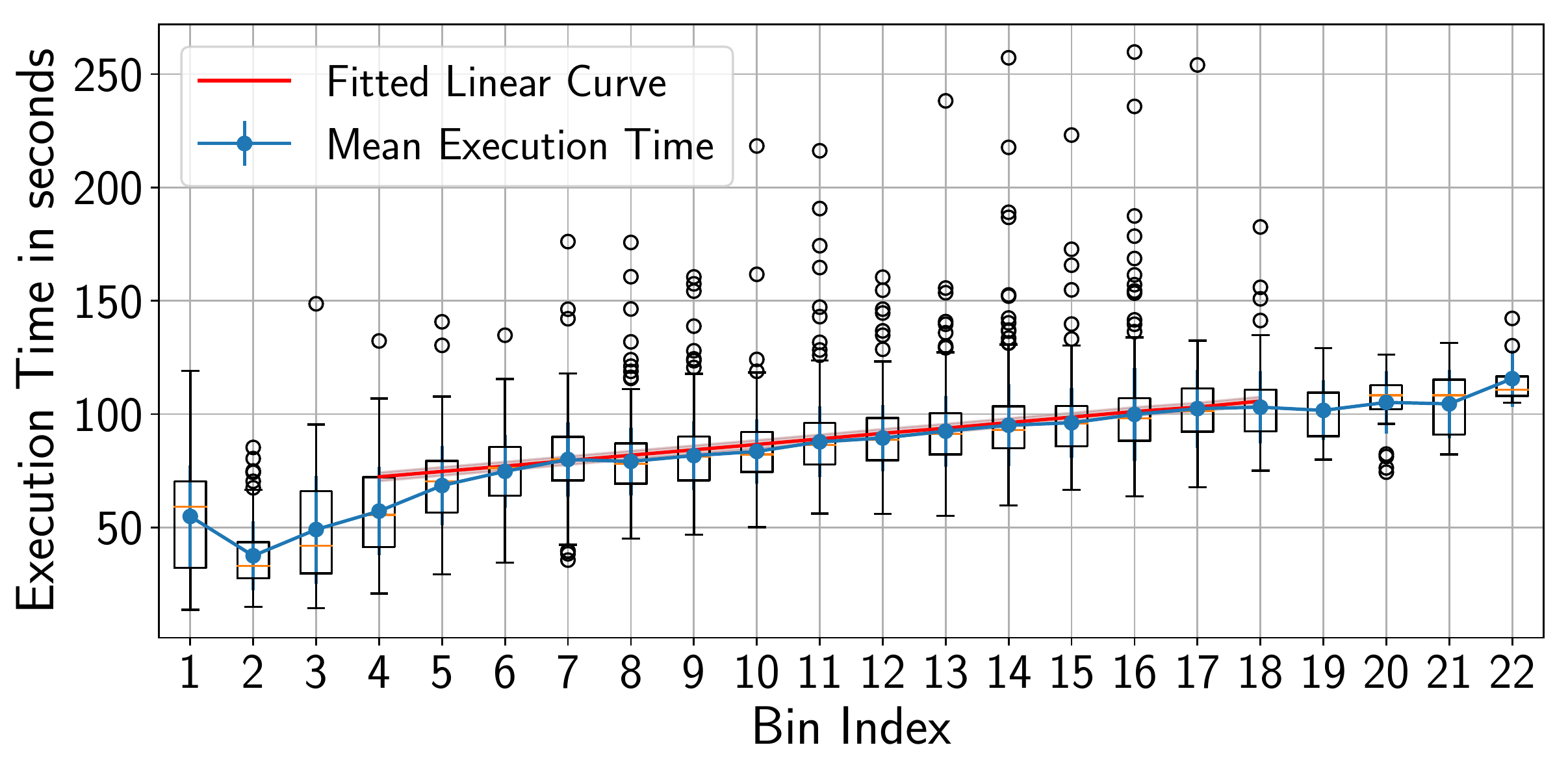}
    \caption{Experiment~1, Design~1: Box-plots of $T_{1}$ execution time,
    mean and $STD$ for $125$~MB image size bins. Red line shows fitted linear
    function.}\label{fig:stage_0_execution}
\end{figure}

Fig.~\ref{fig:stage_0_execution} shows the execution time of the tiling
task---defined as $T_{1}$ in~\S\ref{des1}---as a function of the image size.
We partition the set of images based on image size, obtaining 22 bins of
$125$~MB each in the range of $[50;2,800]$~MB. The average time taken to tile
an image in each bin tends to increase with the size of the image. The
box-plots show some positive skew of the data with a number of data points
falling outside the assumed normal distribution. There are also large
standard deviations ($STD$, blue line) in most of the bins. Thus, there is a
weak correlation between task execution time and image size with a large
spread across all the image sizes.

\begin{figure}[ht!]
    \centering
    \includegraphics[width=.45\textwidth]{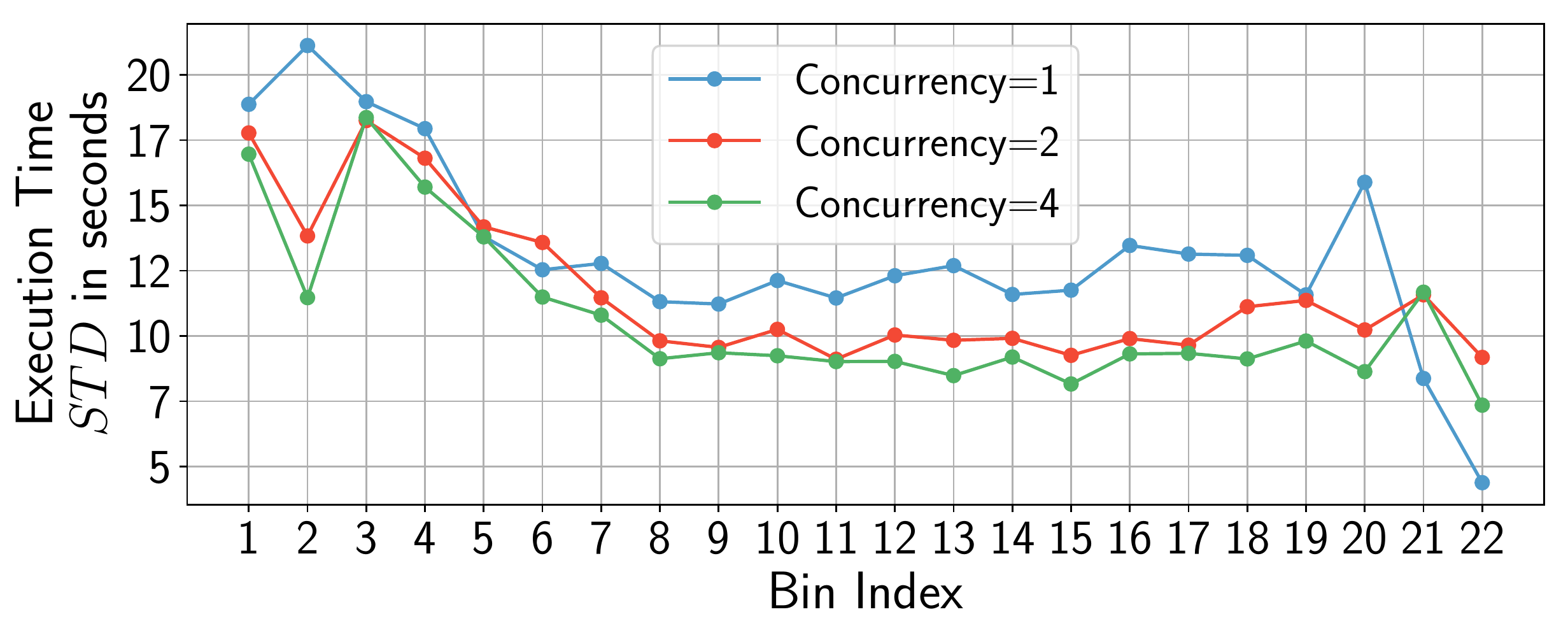}
    \caption{$STD$ of $T_{1}$ execution time based on image size bin and
    number of concurrent tasks. Mostly dependent on compute node's
    performance and invariant across values of task
    concurrency.}\label{fig:concurrency_test}
\end{figure}

We explored the causes of the observed $STD$ by measuring how it varies in
relation to the number of tiling tasks concurrently executing on the same
node. Fig.~\ref{fig:concurrency_test} shows the standard deviation of each
bin of Fig.~\ref{fig:stage_0_execution}, based on the amount of used task
concurrency. We observe that $STD$ drops with increased concurrency but
remains relatively stable between bins $\#6$ and $\#20$. We attribute the
initial dropping to how Lustre's caching improves the performance of an
increasing number of concurrent requests. Further, we observe that as the
type of task and the compute node are the same across all our measures, the
relatively stable and consistent $STD$ observed across degrees of task
concurrency depends on fluctuations in the node performance.

Fig.~\ref{fig:stage_0_execution} indicates that the execution time is a
linear function of the image size between bin \#4 and bin \#18. bins $1-3$
and $19-23$ are not representative as the head and tail of the image sizes
distribution contain less than $5\%$ of the image dataset. We model the
execution time as:
\begin{equation}
	T(x) = \alpha \times x+\beta
    \label{eq:des1_til}
\end{equation} where $x$ is the image size.

We found the parameter values of Eq.~\ref{eq:des1_til} by using a non-linear
least squares algorithm to fit our experimental data, which are $\alpha=
1.92 \times 10^{-2}$, and $\beta = 60.49$ (see red line in
Fig.~\ref{fig:stage_0_execution}). $R^{2}$ of our fitting is $0.97$, showing
a very good fit of the curve to the actual data.

The Standard Error of the estimation, $S_{error}$, reflects the precision of
our regression. The $S_{error}$ is equal to $1.93$, shown as the red shadow
in Fig.~\ref{fig:stage_0_execution}. From $R^{2}$ and $S_{error}$ we conclude
that our estimated function is validated and is a good fit for the execution
time of $T_{1}$ for Design~1.

\begin{figure}[ht!]
    \centering
    \includegraphics[width=0.45\textwidth]{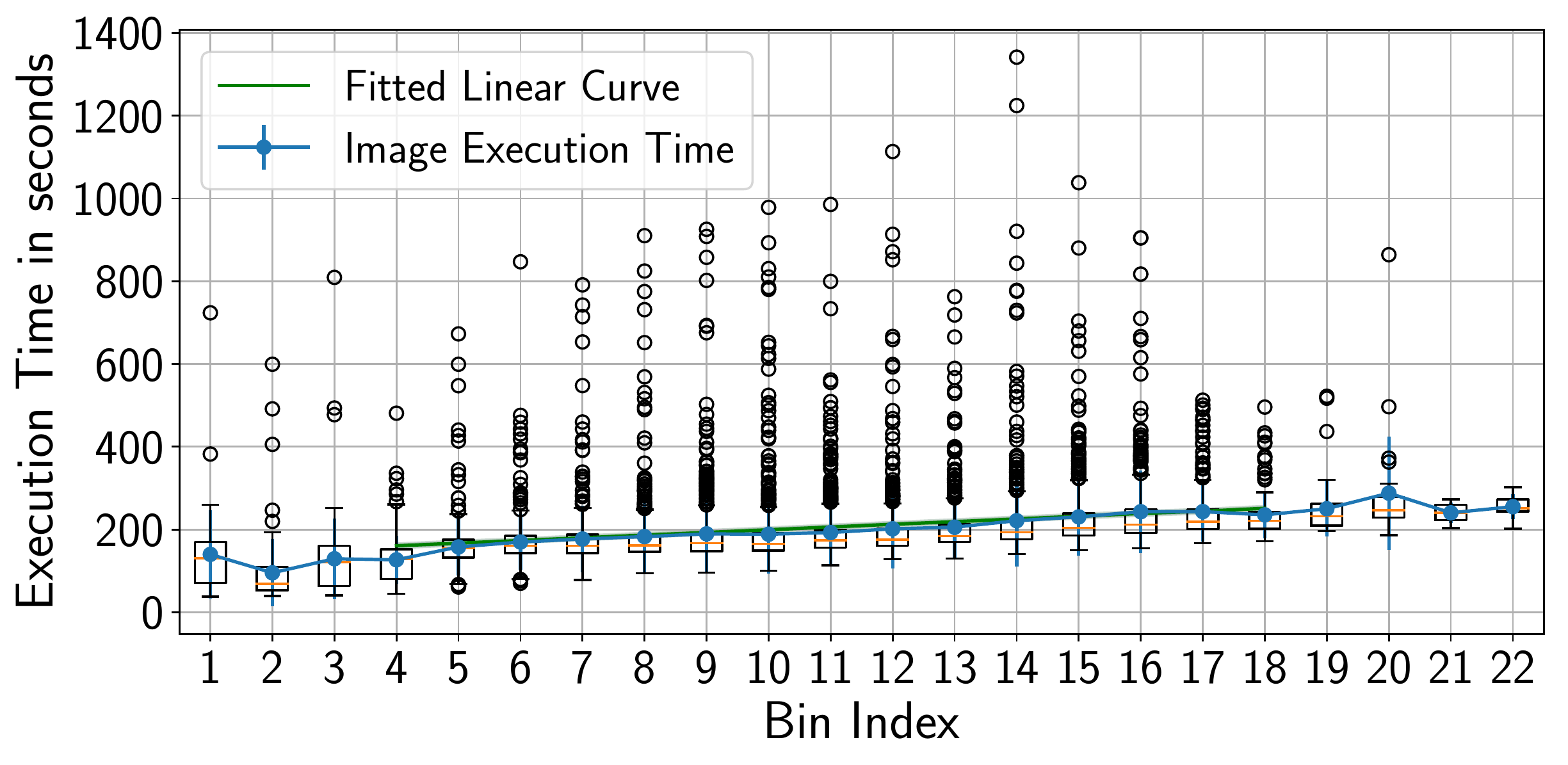}
    \caption{Experiment~1, Design~1: Box-plots of $T_{2}$ execution time,
    mean and $STD$ for $125$~MB image size bins. Green line shows fitted
    linear function.}
    \label{fig:stage_1_execution}
\end{figure}

Fig.~\ref{fig:stage_1_execution} shows the execution time of the seals
counting task as a function of the image size. Defined as $T_{2}$
in~\S\ref{des1}, this task presents a different behavior than $T_{1}$, as the
code executed is different. Note the slightly stronger positive skew of the
data compared to that of Fig.~\ref{fig:stage_0_execution} but still
consistent with our conclusion that deviations are mostly due to fluctuations
in the compute node performance (i.e., different code but similar
fluctuations).

Similar to $T_{1}$, Fig.~\ref{fig:stage_1_execution} shows a weak correlation
between the execution time of $T_{2}$ and image size. In addition, the
variance per bin is relatively similar across bins, as expected based on the
analysis of $T_{1}$. The box-plot and the mean execution time indicate that a
linear function is a good candidate for a model of $T_{2}$. As in
Eq.~\ref{eq:des1_til}, we fitted a linear function to the execution time as a
function of the image size for the same bins as $T_{1}$.

Using the same method we used with $T_{1}$, we produced the green line in
Fig.~\ref{fig:stage_1_execution} with parameter values $\alpha = 5.21
\times 10^{-2}$ and $\beta = 128.53$. $R^{2}$ is $0.96$, showing a good fit
of the line to the actual data, while $S_{error}$ is $5.73$, slightly higher
than for $T_{1}$. As a result, we conclude that our estimated function is
validated and is a good fit for the execution time of $T_{2}$ for Design~1.

\subsection{Experiment~1: Design~2 Tasks Execution Time}

\begin{figure}[ht!]
    \centering
    \includegraphics[width=0.45\textwidth]{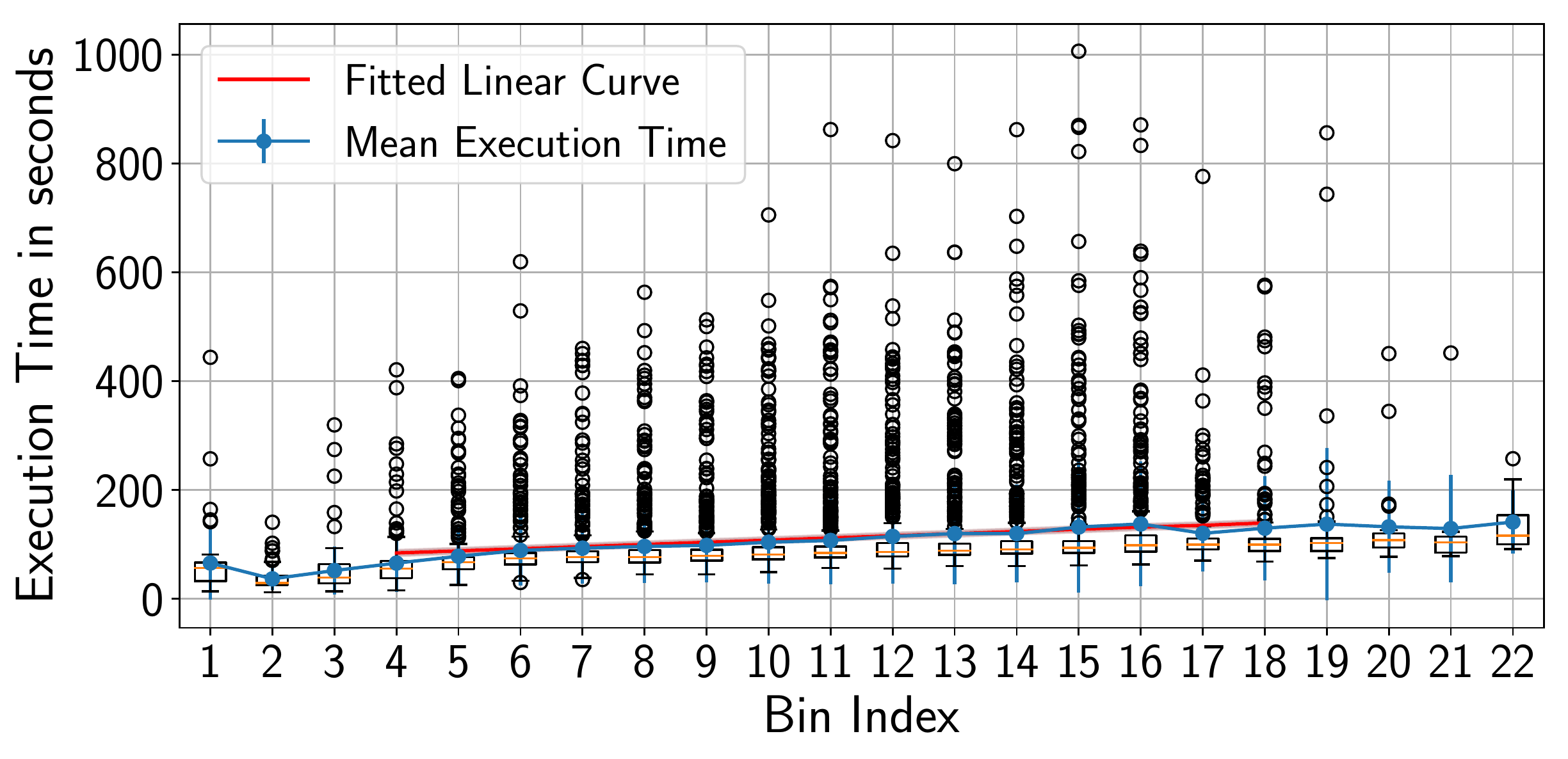}
    \caption{Experiment~1, Design~2: Box-plots of $T_{1}$ execution time,
    mean and $STD$ for $125$~MB image size bins. Red line shows fitted linear
    function.}\label{fig:stage_1_execution_des2}
\end{figure}

Fig.~\ref{fig:stage_1_execution_des2} shows the execution time of $T_{1}$ as
a function of the image size for Design~2. In principle, design differences
in middleware that execute tasks as independent programs should not directly
affect task execution time. In this type of middleware, task code is
independent from that of the middleware: once tasks execute, the middleware
waits for each task to return. Nonetheless, in real scenarios with
concurrency and heterogeneous tasks, the middleware may perform operations on
multiple tasks while waiting for others to return. Accordingly, in Design~2
we observe an execution time variation comparable to that observed with
Design~1 but Fig.~\ref{fig:stage_1_execution_des2} shows a stronger positive
skew of the data in Design~2 than Fig.~\ref{fig:stage_0_execution} in
Design~1.

We investigated the positive skew of the data observed in
Fig.~\ref{fig:stage_1_execution_des2} by comparing the system load of a
compute node when executing the same number of tiling tasks in Design~1 and
2. The system load of Design~2 was higher than that of Design~1. Compute
nodes have the same hardware and operating system, and run the same type and
number of system programs. As we used the same type of task, image and task
concurrency, we conclude that the middleware implementing Design~2 uses more
compute resources than that used for Design~1. Due to concurrency, the
middleware of Design~2 competes for resources with the tasks, momentarily
slowing down their execution. This is consistent with the architectural
differences across the two designs: Design~2 requires resources to manage
queues and data movement while Design~1 has only to schedule and launch tasks
on each node.

We fitted Eq.~\ref{eq:des1_til} to the execution time of $T_1$ for Design~2,
obtaining $\alpha = 3.174 \times 10^{-2}$ and $\beta = 64.81$. The fitting
produced the red line in Fig.~\ref{fig:stage_1_execution_des2}. $R^{2}$ is
$0.92$, showing a good fit of the curve to the data and $S_{error}$ is
$5.50$, validating our estimated function. $R^2$ and especially $S_{error}$
are worse compared to Design~1, an expected difference based on the positive
skew of the data observed in Design~2.

\begin{figure}[ht!]
    \centering
    \includegraphics[width=0.45\textwidth]{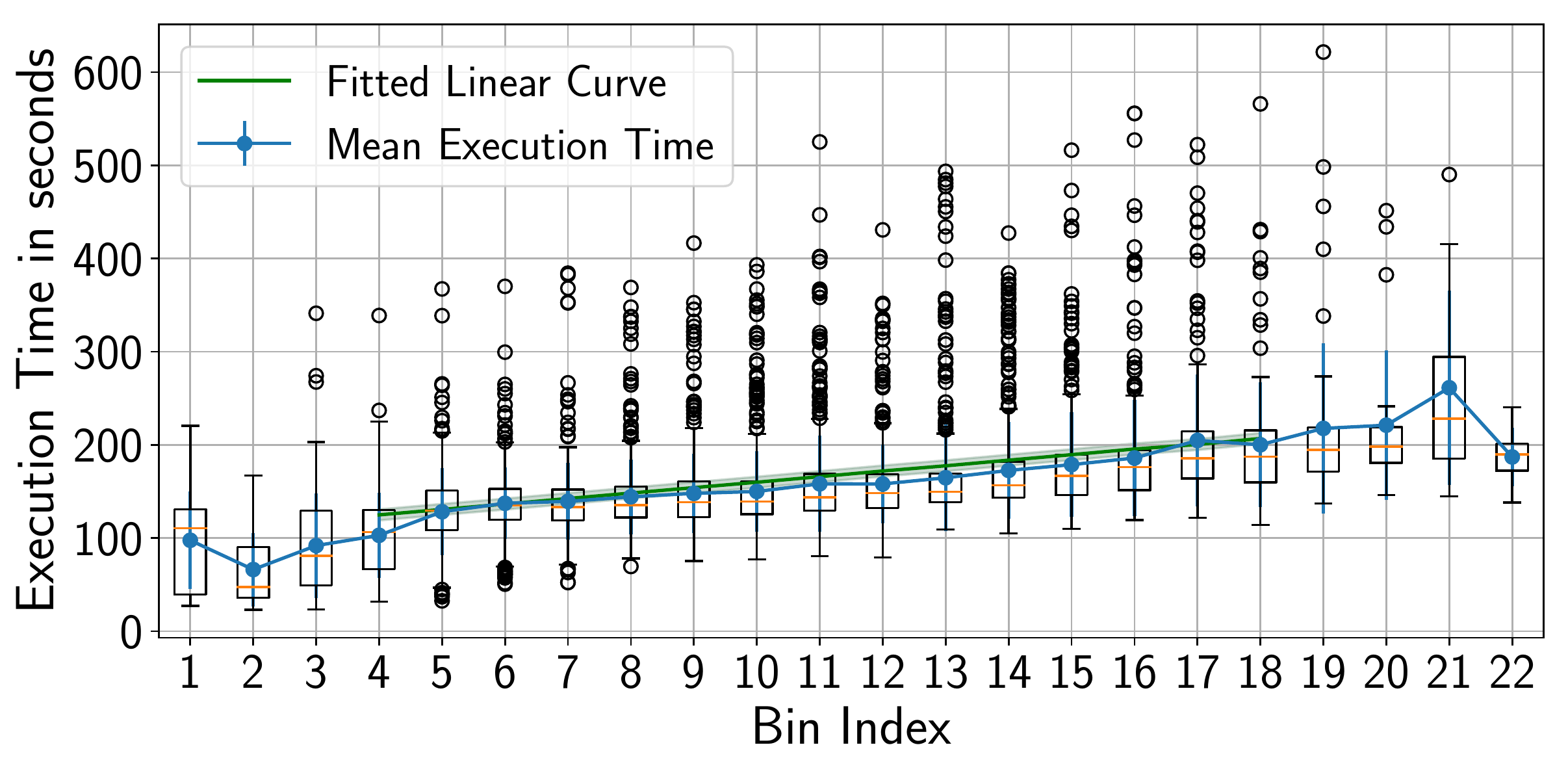}
    \caption{Experiment~1, Design~2: Box-plots of $T_{2}$ execution time,
    mean and $STD$ for $125$~MB image size bins. Green line shows fitted
    linear function.}\label{fig:stage_2_execution_des2}
\end{figure}

Fig.~\ref{fig:stage_2_execution_des2} shows that Design~2 also produces a
much stronger positive skew of $T_{2}$ execution time compared to executing
$T_{2}$ with Design~1. $T_{2}$ executes on GPU and $T_{1}$ on CPU but their
execution times produce comparable skew in Design~2. This further supports
our hypothesis that the long tail of the distribution of $T_{1}$ and
especially $T_{2}$ execution times depends on the competition for main
memory and I/O between the middleware implementing Design~2 and the executing
tasks.

Fitting the model gives $\alpha = 4.71 \times 10^{-2}$ and $\beta = 95.83$.
$R^{2}$ is $0.95$ and $S_{error}$ of $5.96$. As a result, the model is
validated and a good candidate for the experimental data. We attribute the
difference between these values and those of the model of $T_{2}$ for
Design~1 to the already described positive skew of execution times in
Design~2.

\subsubsection{Design~2.A} Similarly to the analysis in Design~1 and 2 we
fitted data from Design~2.A to Eq.~\ref{eq:des1_til}. For $T_{1}$ the fit
gives $\alpha=2.74\times10_{-2}$ and $\beta=49.03$, with $R^{2}$ and
$S_{error}$ of $0.94$ and $3.89$ respectively. For $T_{2}$ the fit gives
$\alpha=4.8\times10_{-2}$ and $\beta=87.36$, with $R^{2}=0.95$ and
$S_{error}=6.19$. Both models are therefore a good fit for the data and are
validated.

\mtnote{The results of our experiments indicate that, on average and across
multiple runs, there is a decrease in the execution time of $T_{1}$ compared
to Design~2 but not of $T_{2}$. Design~2.A requires one queue more than
Design~2 for $T_{1}$ but the execution of $T_{2}$ does not vary between the
two designs. Note that in Design~2.A the image dataset of $T_{1}$ is local to
the node. This explains $T_{1}$ shorter execution time: pulling images for a
local filesystem instead of the Lustre network filesystem outweigh the added
overhead of managing an additional queue.}

The results of our experiments indicate that, on average and across multiple
runs, there is a decrease in the execution time of $T_{1}$ and an increase in
that of $T_{2}$ compared to Design~2. Design~2.A requires one queue more than
Design~2 for $T_{1}$ and therefore more resources for Design~2.A
implementation. This can explain the slowing of $T_{2}$ but not the speedup
of $T_{1}$. This requires further investigation, measuring whether the
performance fluctuations of compute nodes are larger than measured so far.

\begin{figure*}[ht!]
    \centering
    \begin{subfigure}[b]{0.49\textwidth}
        \includegraphics[width=\linewidth]{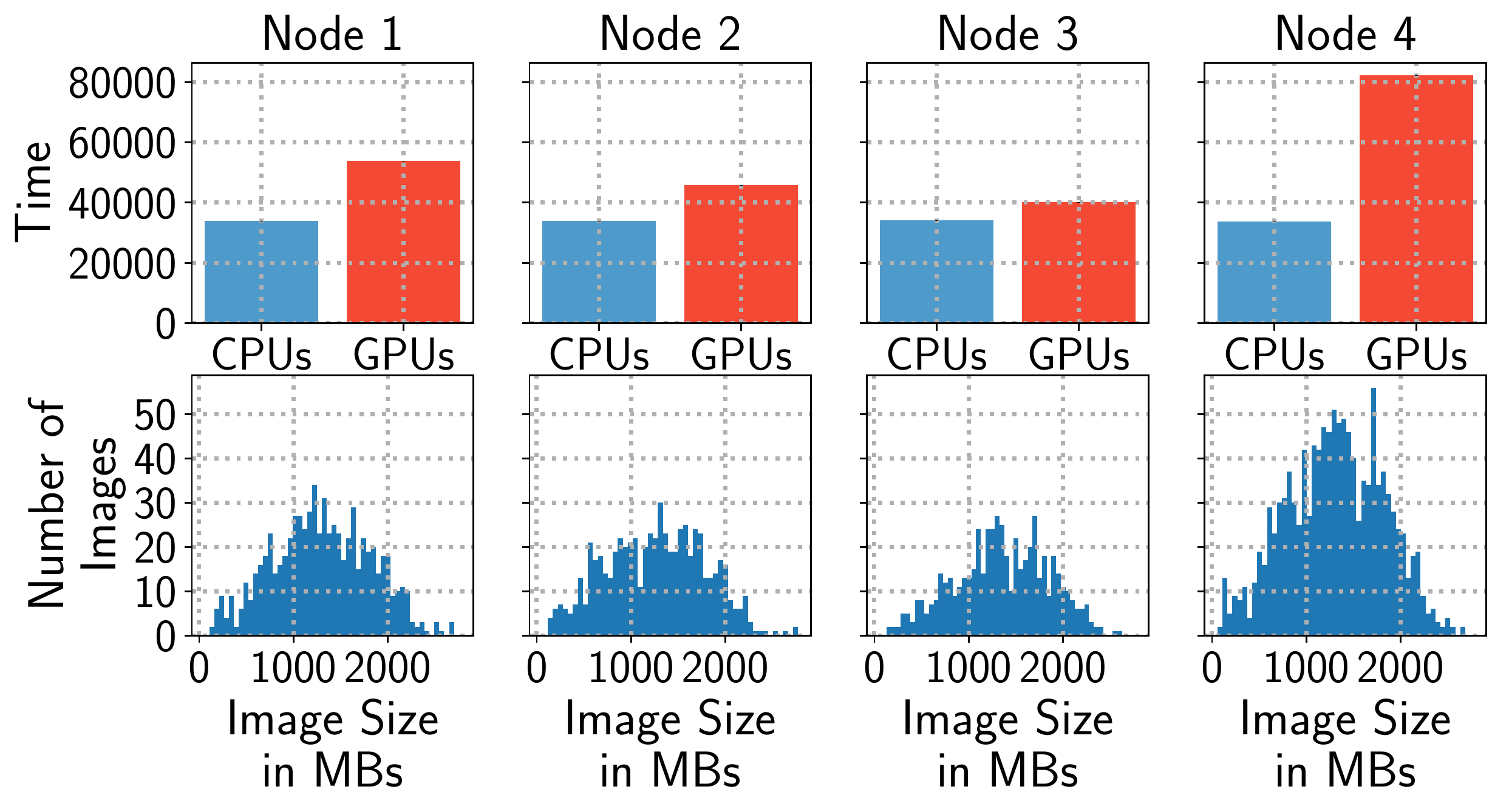}
        \caption{}
        \label{fig:design2_timeline}
    \end{subfigure}%
    ~ 
    \begin{subfigure}[b]{0.49\textwidth}
        \includegraphics[width=\linewidth]{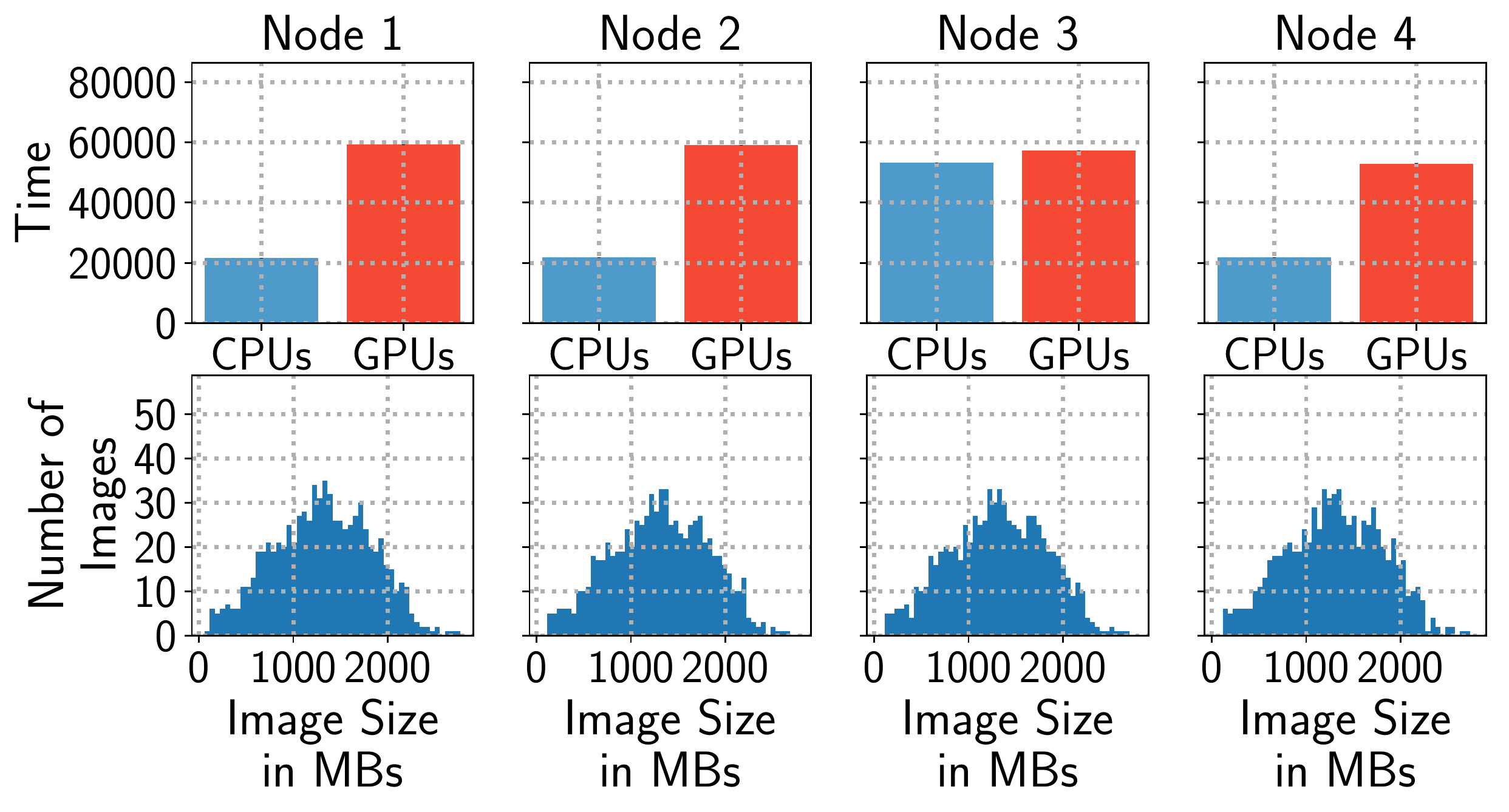}
        \caption{}
        \label{fig:design2a_timeline}
    \end{subfigure}
    \caption{Execution time of $T_{1}$ (blue) and $T_{2}$ (red),
    and distributions of image size per node for (a) Design~2 and (b)
    Design~2.A.}
    \label{fig:design_balancing}
\end{figure*}

As discussed in~\S\ref{ssec:approach2}, balancing of workflow execution
differs between Design~2 and Design~2.A. Fig.~\ref{fig:design2_timeline}
shows that each $T_{1}$ task can work on a different number of images but all
$T_{1}$ tasks concurrently execute for a similar duration. The four
distributions in Fig.~\ref{fig:design2_timeline} also show that this
balancing can result in different input distributions for each compute node,
affecting the total execution time of $T_{2}$ tasks on each node. Thus,
Design~2 can create imbalances in the time to completion of $T_{2}$, as shown
by the red bars in Fig.~\ref{fig:design2_timeline}.

Design~2.A addresses these imbalances by early binding images to compute
nodes. Comparing the lower part of Fig.~\ref{fig:design2_timeline} and
Fig.~\ref{fig:design2a_timeline} shows the difference between the
distributions of image size for each node between Design~2 and 2.A. In
Design~2.A, due to the modeled correlation between time to completion and the
size of the processed image, the similar distribution of the size of the
images bound to each compute node balances the total processing time of the
workflow across multiple nodes.

Note that Fig.~\ref{fig:design_balancing} shows just one of the runs we
perform for this experiment. Due to the random pulling of images from a
global queue performed by Design~2, each run shows different distributions of
image sizes across nodes, leading to large variations in the total execution
time of the workflow. Fig.~\ref{fig:design2a_timeline} shows also an abnormal
behavior of one compute node: For images larger than $1.5$GBs, Node 3 CPU
performance is markedly slower than other nodes when executing $T_{1}$.
Different from Design~2, Design~2.A can balance these fluctuations in $T_{1}$
as far as they don't starve $T_{2}$ tasks.

\subsection{Experiment~2: Resource Utilization Across
Designs}\label{ssec:exp2}

Resource utilization varies across Design~1, 2 and 2.A. In Design~1, the RTS
(RADICAL-Pilot) is responsible for scheduling and executing tasks. $T_{1}$ is
memory intensive and, as a consequence, we were able to concurrently execute
3 $T_{1}$ on each compute node, using only 3 of the 32 available cores. We
were instead able to execute 2 $T_{2}$ concurrently on each node, using all
the available GPUs. Assuming ideal concurrency among the 4 compute nodes we
utilized in our experiments, the theoretical maximum utilization per node
would be $10.6\%$ for CPUs and $100\%$ for GPUs.

\begin{figure*}[ht!]
    \centering
    \begin{subfigure}[b]{0.33\textwidth}
        \includegraphics[width=\linewidth]{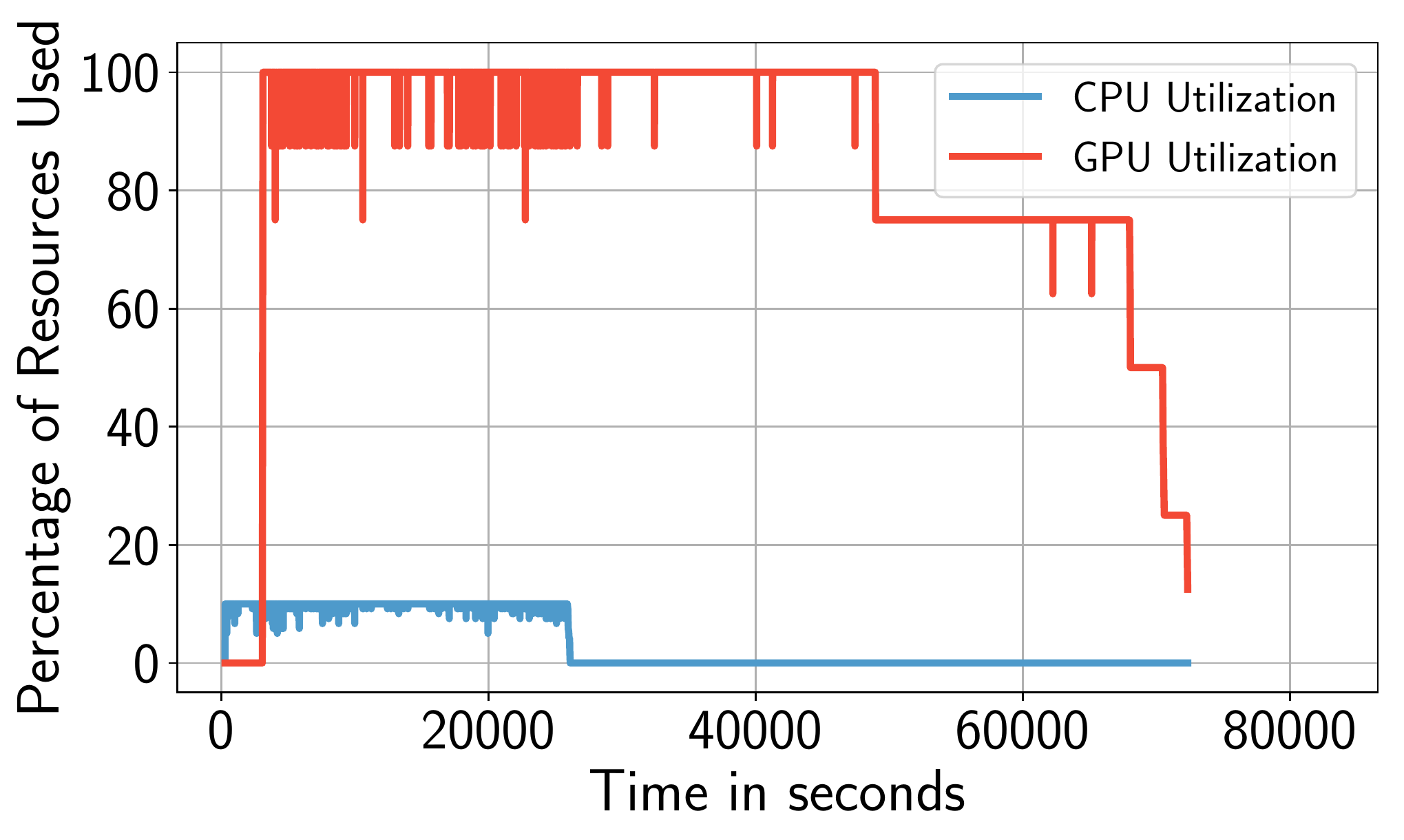}
        \caption{}
        \label{fig:design1util}
    \end{subfigure}%
    ~ 
    \begin{subfigure}[b]{0.33\textwidth}
        \includegraphics[width=\linewidth]{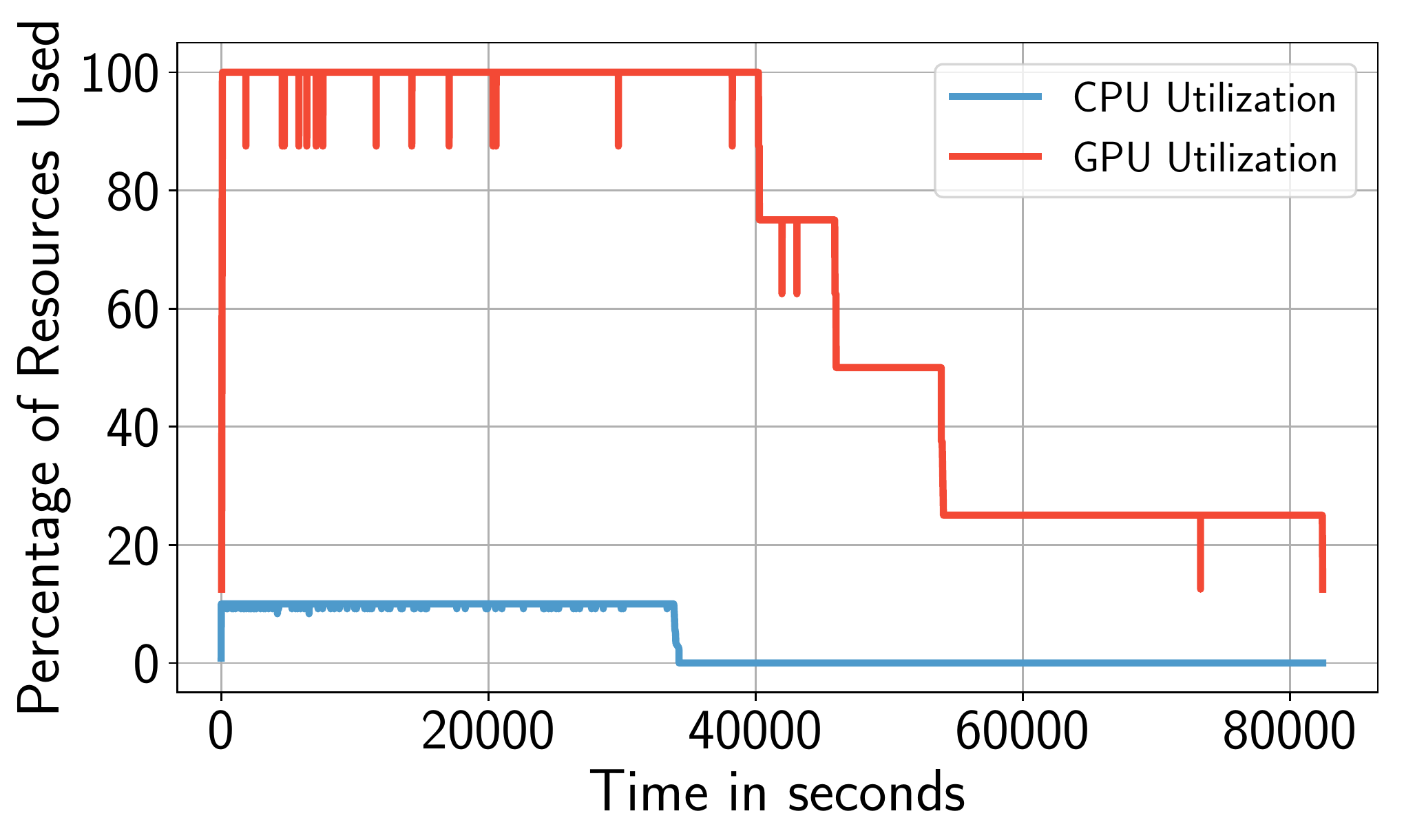}
        \caption{}
        \label{fig:design2util}
    \end{subfigure}%
    ~ 
    \begin{subfigure}[b]{0.33\textwidth}
        \includegraphics[width=\linewidth]{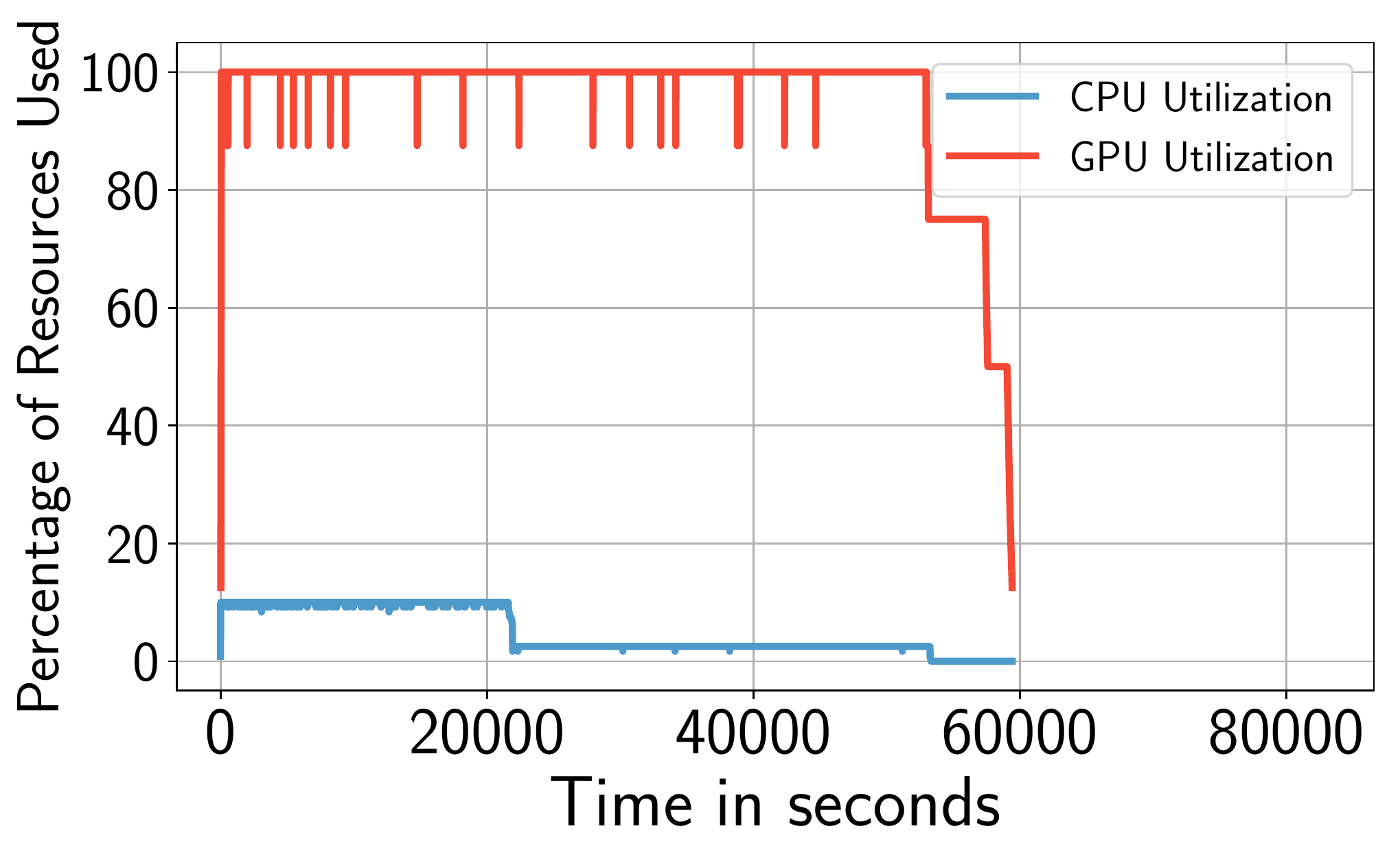}
        \caption{}
        \label{fig:design2autil}
    \end{subfigure}
    \caption{Percentage of CPU and GPU utilization for: (a) Design~1; (b)
    Design~2, and (3) Design~2.A.}
    \label{fig:Utilization}
\end{figure*}

Fig.~\ref{fig:Utilization} shows the actual resource utilization, in percent
of resource type for each design. The actual CPU utilization of Design~1
(Fig.~\ref{fig:design1util}) closely approximates theoretical maximum
utilization but GPU utilization is well below the theoretical $100\%$. GPUs
are not utilized for almost an hour at the beginning of the execution and
utilization decreases to $80\%$ some time after half of the total execution
was completed. 

Analysis shows that RADICAL-Pilot's scheduler did not schedule GPU tasks at
the start of the execution even if GPU resources were available. This points
to an implementation issue and not to an inherent property of Design~1. The
drop in GPU utilization is instead explained by noticing that, as explained
in~\S\ref{sec:design}, GPU tasks where pinned to specific compute nodes so to
avoid I/O bottlenecks. Our experiments confirm that this indeed reduces
utilization as some of the GPU tasks on some nodes take longer time to
process than those on other nodes.

Fig.~\ref{fig:design2util} shows resource utilization for a specific run of
Design~2. GPUs are utilized almost immediately as images are becoming
available in the queues between $T_{1}$ and $T_{2}$, and this quickly leads
to fully utilized resources. CPUs are utilized for more time compared to
Design~1, which is expected due to the longer execution times measured and
explained in Experiment~1. In addition, two GPUs ($25\%$ GPU utilization) are
used for more than $20k$ seconds compared to other GPUs. This shows that the
additional execution time of that node was only due to the data size and not
due to idle resource time.

Fig.~\ref{fig:design2autil} shows the resource utilization for a specific run
of Design~2.A. For 3 compute nodes out of 4, CPU utilization is shorter than
for Design~1 and 2. For the 4th compute node, CPU utilization is much longer
as already explained when discussing $T_{1}$ execution time for Node 3 in
Fig.~\ref{fig:design_balancing}, Experiment~1. As already mentioned, the
anomalous behavior of Node 3 support our hypothesis that compute node
performance fluctuations can be much wider than expected.

Fig.~\ref{fig:design2autil} shows that in Design~2.A GPUs are released faster
compared to Design~1 and Design~2, leading to a GPU utilization above $90\%$.
As already explained in Experiment~1, this is due to differences in data
balancing among designs. This shows the efficacy of two design choices for
the concurrent execution of data-driven, compute-intense and heterogeneous
workflows: early binding of data to node with balanced distribution of image
size alongside the use of local filesystems for data sharing among tasks.

Note that drops in resource utilization are observed in all three designs. In
Design~1, although both CPUs and GPUs were used, in some cases CPU
utilization dropped to 6 cores. Our analysis showed that this happened when
RADICAL-Pilot scheduled both CPU and GPU tasks, pointing to an inefficiency
in the scheduler implementation. Design~2 and 2.A CPU utilization drops
mostly by one CPU where multiple tiling tasks try to pull from the queue at
the same time. This confirm our conclusions in Experiment~1 about resource
competition between middleware and executing tasks. In all designs, there is
no significant fluctuations in GPU utilization, although there are more often
in Design~1 when CPU and GPUs are used concurrently.

\subsection{Experiment~3: Designs Implementation Overheads}\label{ssec:exp3}

\begin{figure}[ht!]
    \centering
    \begin{subfigure}{0.45\textwidth}
    \includegraphics[width=\linewidth]{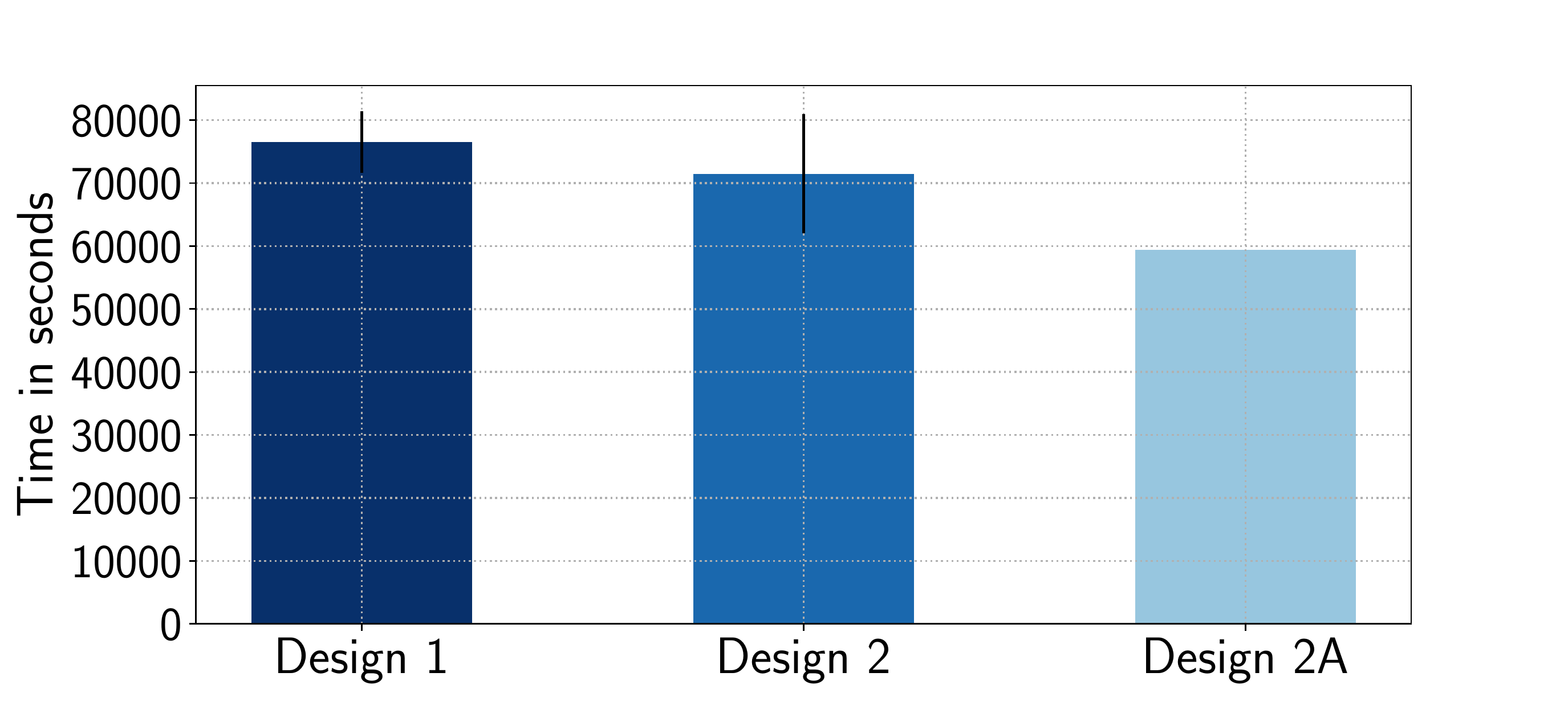}
    \caption{}
    \label{fig:ttx}
    \end{subfigure}\\
    ~
    \begin{subfigure}{0.45\textwidth}
    \includegraphics[width=\linewidth]{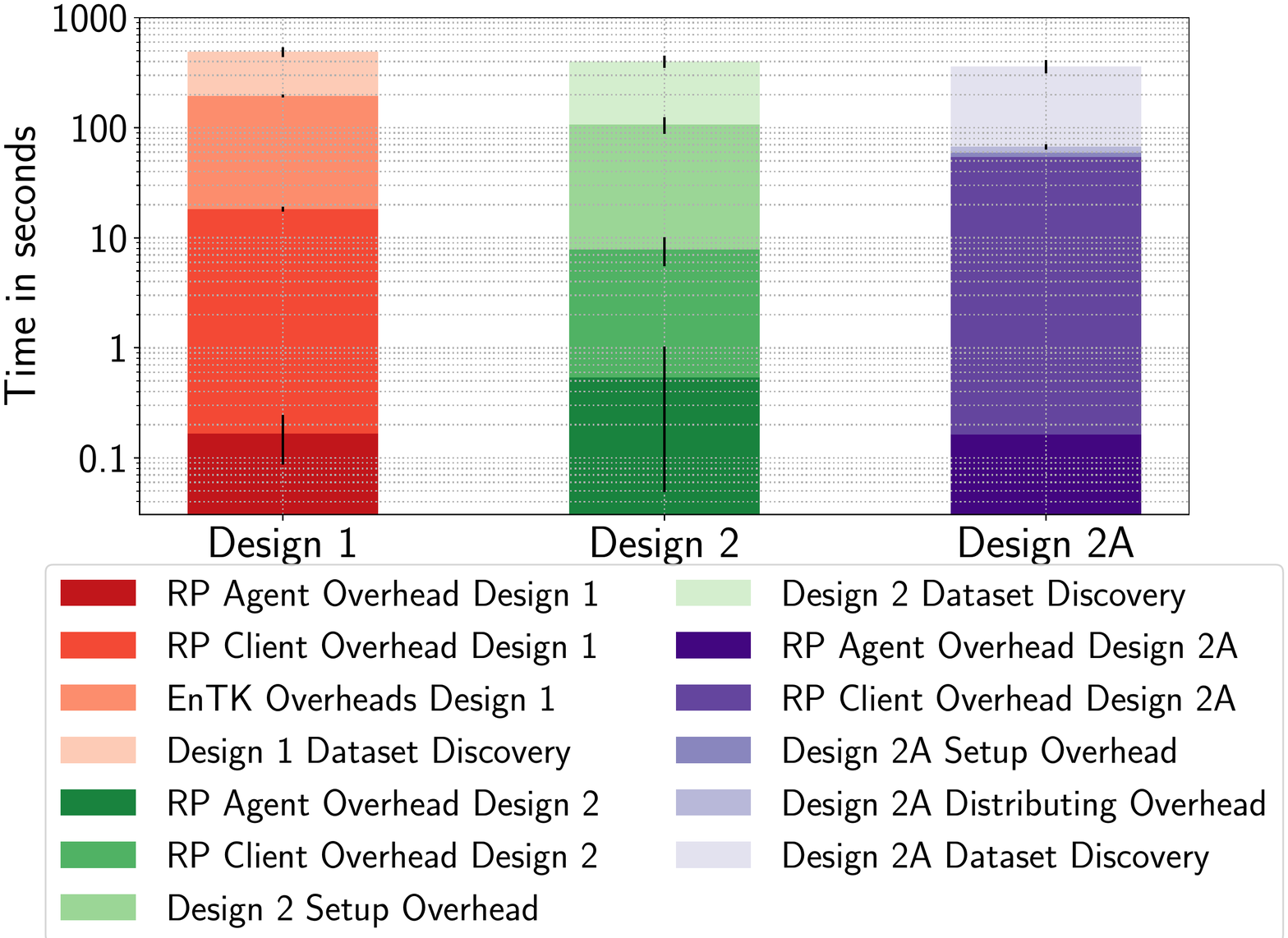}
    \caption{}
    \label{fig:overheads}
    \end{subfigure}
    \caption{(a) Total execution time of Design~1, 2 and 2.A. Design~1 and 2
    have similar performance, Design~2.A is the fastest. (b) Overheads of
    Design~1, 2 and 2.A are at least two orders of magnitude less than the
    total execution time.}\label{fig:overall_performance}
\end{figure}

This experiment studies how the total execution time of our use case workflow
varies across Design~1, 2 and 2.A. Fig.~\ref{fig:ttx} shows that Design~1 and
2 have similar total time to execution within error bars, while Design~2.A
improves on Design~2 and both are substantially faster than Design~1. The
discussion of Experiment~1 and 2 results explains how these differences
relate to the differences in the execution time of $T_{1}$ and $T_{2}$ tasks,
and execution concurrency respectively.

Fig.~\ref{fig:overheads} shows the overheads of each design implementation.
All three designs overheads are at least two orders of magnitude smaller than
the total time to execution. A common overheads between all designs is the
``Dataset Discovery Overhead''. This overhead is the time needed to list the
dataset and it is proportional to the size of the dataset. RADICAL-Pilot has
two main components: Agent and Client. RADICAL-Pilot Agent's overhead is less
than a second in all designs while RADICAL-Pilot Client's overhead is in the
order of seconds for all three designs. The latter overhead is proportional
to the number of tasks submitted simultaneously to RADICAL-Pilot Agent.

EnTK's overhead in Design~1 includes the time to: (1) create the workflow
consisting of $3.1k$ independent pipelines; (2) start \entk's components; and
(3) submit the tasks that are ready to be executed to RADICAL-Pilot. This
overhead is proportional to the number of tasks in the first stage of a
pipeline, and the number of pipelines in the workflow. EnTK does not
currently support partial workflow submission, which would allow us to submit
the minimum number of tasks to fully utilize the resources before submitting
the rest. Experiments should be performed to measure the offset between
resource utilization optimization and increased time spent in communicating
between EnTK and RADICAL-Pilot.

Fig.~\ref{fig:overheads} shows that the dominant overheads of Design~2 is
``Design~2 Setup Overhead''. This overhead includes setting up and starting
queues in each compute node, and starting and shutting down both $T_{1}$ and
$T_{2}$ tasks on each node. Setting up and starting the queues accounts for
most of the overhead as we use a conservative waiting time to assure that all
the queues are indeed up and ready. This can be optimized further reducing
the impact of this overhead.

Alongside the overheads already discussed, Design~2.A also introduces an
overhead called ``Design~2.A Distributing Overhead'' when partitioning and
distributing the dataset over separate nodes. The average time of this
overhead is $7.5$ seconds, with a standard deviation of $3.71$ and is
proportional to the dataset and the number of available compute nodes.

In general, Design~2.A shows the best and more stable performance, in terms
of overheads, resource utilization, load balancing and total time to
execution. Although Design~2 has similar overheads, even assuming
minimization of Setup Overhead, it does not guarantee load balancing as done
instead by 2.A. Design 1 separates the execution in independent pipelines
that are independently executed by the runtime system on any available
resource. Based on the results of our analysis, both EnTK and RADICAL-Pilot
can be configured to implement early binding of images to each compute node
as done in Design~2.A. Nonetheless, Design~1 would still require executing a
task for each image, imposing bootstrap and tear down overheads for each
task.

\section{Discussion and Conclusion}\label{sec:conclusion}

While design 1, 2 and 2.A can successfully support the execution of the
paradigmatic use case described in~\S\ref{sec:ucase}, our experiments show
that for the metrics considered, Design~2.A is the one that offers the better
overall performance. Generalizing this result, use cases that are both
data-driven and compute-intense benefit from early binding of data to compute
node so to maximize data and compute affinity and equally balance input 
across nodes. This approach minimizes the overall time to completion of this 
type of workflows while maximizing resource utilization.

Our analysis also shows the limits of an approach where pipelines, i.e.,
compute tasks, are late bound to compute nodes. In program-based designs, the
overhead of bootstrapping programs needs to be minimized insuring that each
pipeline processes as much input as possible (in our use case, images). In
presence of large amount of data, late binding implies copying, replicating
or accessing data over network and at runtime. We showed that, in
contemporary HPC infrastructures, this is too costly both for resource
utilization and total time to completion.

Infrastructure-wise, our experiments show the limits imposed by an unbalance
between number of CPU cores and available memory. Given data-driven
computation where multi GB images need processing, we were able to use just
$10\%$ of the available cores due to the amount of RAM required. This applies
also to the unbalance between CPUs and GPUs: use cases with heterogeneous
tasks would benefit from a n higher GPU/CPU ratio.

These results apply to the evaluation of the design of existing and future
computing frameworks. For example, we will be extending both \entk{} and
RADICAL-Pilot to implement Design~2.A. We will use our overheads
characterization as a baseline to evaluate our production-grade
implementation and further improve the efficiency of our middleware. Further,
we will apply the presented experimental methodology to other use cases and
infrastructures, measuring the trade offs imposed by other types of task
heterogeneity, including multi-core or multi-GPU tasks that extends beyond a
single compute node.

Beyond design, methodological and implementation insights, the work done with
for this paper has already enabled the execution of our use case at
unprecedented scale and speed. The $3,097$ images of our dataset can be
analyzed in less than 20 hours, compared to labor-intensive weeks previously
required on non-HPC resources.

\footnotesize{{\it Acknowledgements} We thank Andre Merzky (Rutgers) and Brad
  Spitzbart (Stony Brook) for useful discussions.  This work is funded by NSF
  EarthCube Award Number 1740572.  Computational resources were provided by NSF
  XRAC awards TG-MCB090174. We thank the PSC Bridges PI and Support Staff for
  supporting this work through resource reservations.}

\footnotesize{\textit{Software and Data Source Scripts}: \url{http://github.com/iceberg-project/Seals/}, 
    \textit{Experiments and Data}:\url{http://github.com/radical-experiments/iceberg_seals/}}
\bibliographystyle{unsrt}
\bibliography{local}

\end{document}